# Security implications of converged networks and protecting them, without compromising efficiency

By

**Saltuk Aksahin**

A DISSERTATION

Submitted to

The University of Liverpool

in partial fulfillment of the requirements
for the degree of

MASTER OF SCIENCE

**11 February 2007**

# ABSTRACT

## SECURITY IMPLICATIONS OF CONVERGED NETWORKS AND PROTECTING THEM, WITHOUT COMPROMISING EFFICIENCY

By

**Saltuk Aksahin**

This dissertation has extensively looked into all aspects of VoIP communications technology, and information presented in preceding chapters, which build up a solid framework to discuss the conceptual design model, and investigate features that could be incorporated for actual Projects, with parameters that are tested on field values. The dissertation follows a five-course model, for answering different questions, both technical and businesslike, around central issues, that have been crucial to explanation of the topic; starting with a general overview of VoIP technology, analyzing current VoIP encryption methods, identifying security threats, designing a robust VoIP system based on particulars discussed in preceding chapters, and finally, a VoIP simulation.

## CERTIFICATION STATEMENT

I hereby certify that this dissertation constitutes my own product, that where the language of others is set forth, quotation marks so indicate, and that appropriate credit is given where I have used the language, ideas, expressions or writings of another.

                                                February 11, 2007
                                                Doha – Qatar
                                                Saltuk Aksahin

# ACKNOWLEDGEMENTS

I would like to thank my family for supporting me throughout the months this study lasted.

I also would like to thank SKYORYX JV (TAV – TAISEI) MEP team members' for their support and patience.

# TABLE OF CONTENTS

# LIST OF TABLES

# LIST OF FIGURES

## Chapter 1. INTRODUCTION

Voice over Internet Protocol (VoIP) is the communications technology that enables users to route voice conversations over the Internet, or any other IP network, through established network protocols, that bear semblance to ARPANET, and the infrastructure needed is conventional fiber optic data cable. In general, phone services via VoIP come out to be cheaper than Public Standard Telephone Network (PSTN), because cost savings are recorded due to *convergence*, wherein voice and data are carried through a single network. VoIP to VoIP calls are generally free, whereas VoIP to PSTN phone calls bear nominal charges (PCWorld, 2005).

There are 3 principal ways of fixing VOIP technology in order to make phone calls (VoIPinfo, 2006):

- <u>Using an ATA</u>: An Analog Telephone Adaptor (ATA) connects an analog telephone to a VoIP network, through an Ethernet Jack. These are sometimes referred to, as VoIP Gateways. They normally use SIP protocol.
- <u>Using an IP Phone</u>: A broadband hard phone is a self-contained IP telephone, that looks similar to a conventional telephone, but instead of a conventional phone jack, it has an Ethernet port, through which it communicates directly into a VoIP server.
- <u>VoIP connecting directly</u>: It is also possible to *bypass* VoIP service provider. This is the route map for such an arrangement.

IP Phone - Ethernet - Router - Internet - Router - Ethernet - IP Phone

## 1.1 Advantages of VoIP

There are 2 principal advantages of using VoIP service, which explains its popularity in restricted bandwidth telephony (VoIPinfo, 2006):

- Lower costs: As discussed earlier, VoIP to VoIP phone calls are free. E.g. FreeWorldDialUp, Skype and Globe7 are popular services to this effect. Call center agents are the biggest beneficiary of VoIP services
- Increased functionality and mobility: Using VoIP, it is possible to automatically route international calls to a VoIP phone. VoIP phones can be carried globally without paying *roaming* charges. Alternatively, using instant messenger-based services like Skype, VoIP calls can be made from *any* computer as long as there is an Internet connection. Many VoIP services are coming with useful features, such as 3-way calling, call forwarding, automatic redial and caller ID.

## 1.2 Limitations of VoIP services

(The main aim of this Project)? These are some of the problems encountered with VOIP services. Wherever possible, solutions are suggested (VoIPinfo, 2006):

- Fax problems: Currently, not much is been developed commercially for sending Faxes using VoIP. A fax protocol called T.38 is underway.
- Dependence on Internet connections: VoIP services quality varies from country to country, because of its reliance on parameters like Internet speed, bandwidth, etc. These are also subject to power outrages, unlike conventional phones, although UPS-based systems can take care of this problem, somehow.

- <u>Security and transmission difficulties</u>: Many VoIP services suffer due to delay in transmission of voice packets (*Latency*), there is echo, jitter, packet loss and other Quality of Service (*QoS*) problems associated with VoIP phones. As a majority of VoIP users, still do not subscribe to *encryption* services, VoIP poses security (eavesdropping) hazards. A substantial part of this paper is dedicated to studying this phenomena, and the response of data transmission companies.

- <u>Miscellaneous</u>: Inadequate caller ID functions (sometimes, VoIP services cannot trace calls), no emergency calls provision, no free calls to subscriber, all these problems reflect that VoIP is still in its infancy, and there is sometime before it can pick up.

The aim in this Project, is to investigate the current security implications of VoIP, and to run VoIP in a safe, secure and reliable manner. The main objectives of the Project, have been classified, as follows:

- Overview of VoIP Encryption
- Analysis of current VoIP encryption methods, and their effects to the efficiency of VoIP systems (in transmission problems, *latency*, *jitter*, etc.).
- Identification of risks and threats in a VoIP system where encryption has been allowed.

- Design of a method/methodology for proper and reliable VoIP environment:
- Simulation of proposed technique and analysis of simulated system (Case Study using a standard VoIP service)

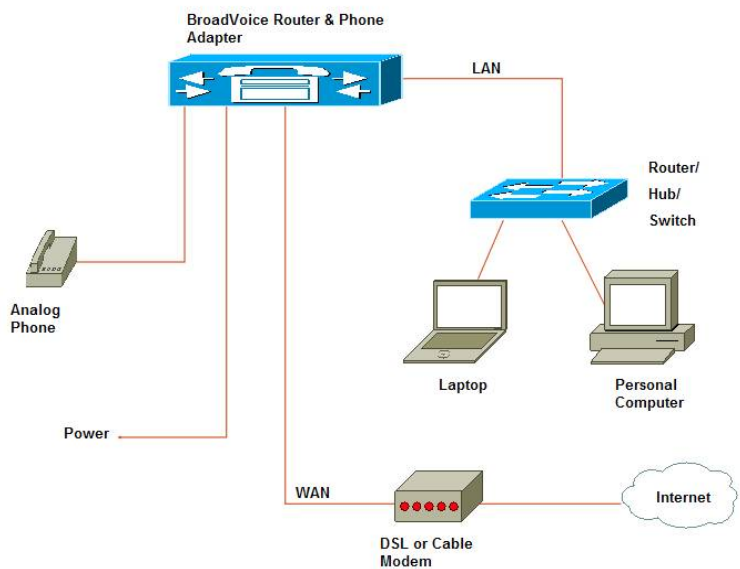

**Figure 1: VoIP System in General**

The main aim of this project is to investigate the current security implications of VOIP and propose a method to run VOIP in a secure way.

Research objective: To look into different aspects of Voice over Internet Protocol (VoIP) communications, mainly Security and Quality issues, and deduct pertinent data for using in a real-time VoIP deployment project. The project aims are 6-fold:
- An overview of VoIP technology

- Identification of common security threats, and failure mechanisms
- Analyzing current VoIP encryption methods, and their efficacy
- Identifying problems with VoIP protocols, e.g. SIP
- Design Methodology
- Project Simulation

<u>Content/Methodology</u>: In order to answer above hypotheses, and based on stated security requirements mandated from experts agreement over this issue, around 13 security ideas have been short listed, and the dissertation has proceeded alongside to take into account any key data, that may be needed in developing the answers to the security issues. One common theme overriding security parameters, is Quality of Service (QoS) values, like *jitter*, *latency*, and *packet loss*. In order to solve for these values, the dissertation has expanded itself into Project Design, and Project Simulation, which uses proprietary software, like Visual Route, Minitab and VoIP Calculator.

<u>Result/Conclusions</u>: VoIP, as proved in this dissertation, has a fascinating development path ahead, and as per stated assumptions, we are able to develop a realistic Design framework, for this technology area, and verify our design parameters for QoS, successfully through Project. All calculations have been shown in Appendices, and the central objective of dissertation has been satisfied many times over.

# Chapter 2.  LITERATURE SURVEY

## 2.1  VOIP Security Basics

According to the Telecommunications Industry Association, Arlington, VA, at some point in 2006, more than 50% of all new private branch exchanges being installed will be IP-based. The number of residential VoIP subscribers, will grow 12-fold to about 12 million (Cherry, p.1). Cherry (2005, p.6) asserts that in this era of speedy telecommunications, bandwidth is no longer an issue, as there is plenty of space for data and voice to be dispatched together, and that the two main things to worry about in VoIP calls, are: *Latency* (data packets are delivered too slowly, due to network congestion), *Jitter* (variation in delay of packets, leading to packet loss) (Cherry, p.8), and thus, having bandwidth is not the only solution for safe, secure and speedy telephony, as the real *fault* lies in what is known as the *last mile* (Cherry, p.7), the connection between home PC's Ethernet port, and fiber optic cable. Even 500 kbps broadband connections can deliver high-quality PC-to-phone telephony (Cherry, p.8), provided the concept of voice delivery is clarified. It must be remembered that VoIP encryption, and other security discussions are interrelated with transmission factors, and cannot be studied in isolation.

The typical VoIP encryption system utilizes public-key cryptography (PKC). Skype, e.g. uses Advanced Encryption Standard, with encryption keys that are 256-bits long. Users log in into their personal Skype accounts, and are then recognized by a Skype server, across the network. The server gives each party a key (Read Skype password) to enable decrypting the encryption key, which enables for a high-end security (Cherry, p.10).

What it basically means, is that VoIP systems are broadly governed by the method of encryption, although theoretically, it is possible to *eavesdrop* on VoIP-conversations anywhere in the network, on a practical scale it must take a hacker (unless he knows password), $2^{256}$ combinations to try before he can hack Skype. That is as safe as Online banking, and credit card transactions on the Internet!

*So, what are the key security issues for VoIP*? According to ideas furnished at Computer Society (ComSoc) seminar on this topic, experts have agreed on the following key issues, worth addressing in this dissertation (ComSoc, 2005):

- Enterprise and carrier network architectures for secured VoIP
- Security issues in p2p VoIP systems
- Cross-layer security architecture
- Carrier Peering and Session Border Control
- Security issues with NAT traversal mechanisms
- Vulnerabilities in VoIP protocols, such as SIP
- End-user and hop-by-hop authentication techniques
- QoS impacts due to security implications (latency, jitter, etc., covered in our dissertation case study)
- SPAM and DOS attacks
- Securing interconnections with SS7 networks
- Securing Voice over WLAN (VoWLAN)

Although not exhaustive, answering above security issues can enable us to develop a robust, and mature VoIP security design, for this Project purpose. The scope of this stream of study, stretches the furthest advancement in R&D related activities. Therefore, we must restrict our design to this effect only. The above security issues address a majority of practical problems, that are faced due to VoIP security hazards.

A little more on Advanced Encryption Standards (AES) before we proceed further. According to the Computer Security Resource Clearinghouse (CSRC)'s declassified report on this topic, AES is defined as a FIPS-approved cryptographic algorithm, that can be used to protect voice data. Encryption basically converts data into an unintelligible form called ciphertext; decrypting the ciphertext back into what is called plain text at the receiver's end (CSRC, p.1). Before initializing any AES software bundle into VoIP installation, following objectives must be kept in mind:

An approving authority and a maintenance agency: Generally, the US Secretary of Commerce is the approving authority, the Department of Commerce and the National Institute of Standards and Technology are the maintenance agencies.

Specifications: Federal Information Processing Standards (FIPS), 197, Advanced Encryption Standard (AES), (affixed).

Implementations: The algorithm specified in this standard may be implemented in software, hardware, firmware, or any combination thereof.

- Where to obtain copies: http://csrc.nist.gov/publications/ after e-payment

## 2.2 Enterprise and carrier network architecture for secure VoIP

A safe and secure network architecture, must have the following components: 1)Router devices 2)Switch devices 3)Load balancing devices 4)Firewall devices 5)Virtual Private Networks (VPN's) (Microsoft.com, 2004)

The first thing a VoIP service provider, would need to ensure, is what is known as LAN segment architecture definition, which forms the principal infrastructure layer for any Project on VoIP service. The network architecture's prime role, is to combine a number of technological solutions, into a complex environment that performs, a highly secure, available, scalable and reliable service. Enterprise-level network architectures generally consist of a holistic mix of any of the following technologies, and elements:

LAN segments: Connectivity speeds between hosts and devices is typically, 10 MBPS or more.

LAN connectivity devices: Switches, routers, load balancers, etc.

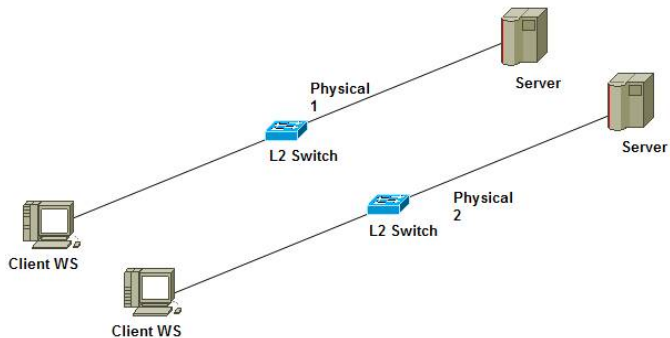

**Figure 2: Simple Physical connection between LAN networks**

Refer fig.2, which shows a real, physical network. In a similar way, we can have a virtual network on L2 and L3 switches, which improves upon transmission speeds greatly, and must be strongly encouraged. Next in stage, is a brief, yet implicit understanding of the 5 parameters mentioned in VoIP basic infrastructure:

Router devices: Switching and routing devices provide efficient movements of packets, from one device to another, at L2 terminal of the OSI model. Devices that perform the L2 switching role, can learn destination addresses, by monitoring network traffic. Many enterprise-level switches, have the capability to route from segments directly. A number of permutations and combinations is possible, but the best router devices, are selected after enough deliberation in optimum design needs.

Load balancing: Load balancing is an activity, that facilitates horizontal clustering, where multiple servers are configured to perform the same role on the network. The load balancing functionality can be provided by a software in any two ways: 1)Distributed (Each node in a cluster, by a specified algorithm receives a packet destined for the cluster), and 2)Routed (A centralized device which receives every inbound packet destined for a cluster, and decides which host to send the packet to).

Firewall devices: The firewall services control the flow of traffic, between network segments, using L3 addresses, in order to meet security requirements. Most firewalls come equipped with an implicit "deny" rule, which means if a request does not exist, the request is automatically denied. Refer Table 1 for more on that.

| Source IP Address | Destination IP Address | Action | Port | Protocol |
|---|---|---|---|---|
| Any | 208.217.184.6 | Allow | 80 | TCP |
| Any | 10.1.0.5 | Deny | 25 | TCP |

**Table 1 : An example of how a Firewall allows or denies, a Destination IP address, from accessing data**

Virtual Private Networks (VPN's): Digital Modems, WAN's, WPAN, Wireless are some examples of private networks, that can be supported on an Enterprise, and Carrier network architecture.

Before laying VoIP infrastructure, all service companies make efforts to define security architecture. It starts with a Security Zone definition, where it is specified, which tier, and which security zone, must the infrastructure come equipped with.

It follows with network role definition, and finally, design of the entire arrangement. There are plenty of design options available. Fig. 3 is one example.

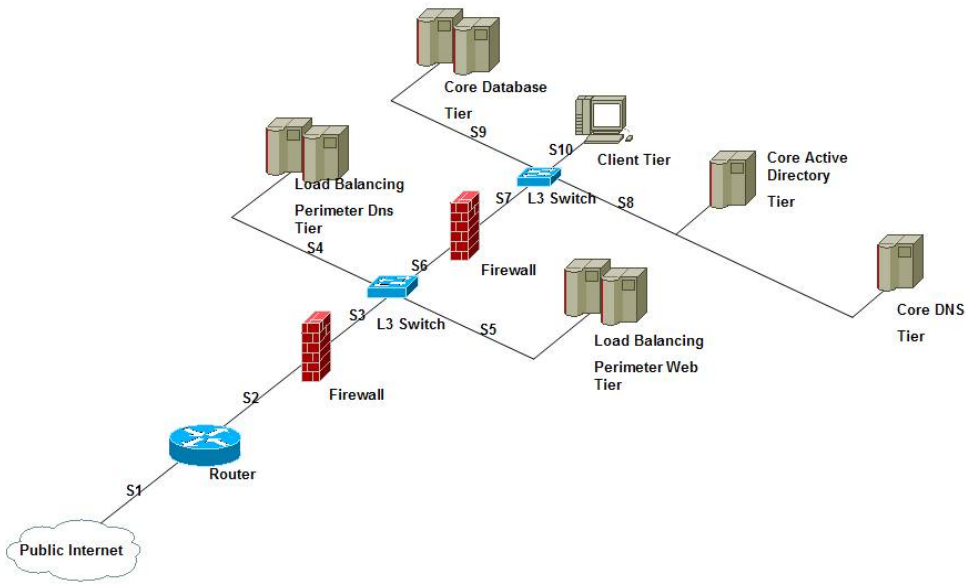

**Figure 3: Enterprise and Carrier Network Architecture design example**

Next-generation network architecture goes a step further in improving upon security issues (MSForum, 2003):

Call agent/SIP Server/SIP Client: This provides call logic, and call control functions, e.g. Caller ID, Call Waiting and interaction with application servers (p.7)

Service broker and application server: Service brokers are present at the edge of service provider end, and allow for a consistency in coordinating various applications. Application server is located in network for same (p.8)

Media server: Sophisticated roles such as playing announcements, tone adjustment, interactive voice response processing, etc. (p.8)

Gateways: Signaling, trunking and access gateway (p.9)

Bandwidth manager: It provides required QoS for the network. (p.9)

Edge router: It is located in service provider's network, and is mainly used to improve packet flow at ends. (p.9)

There are 3 security issues that a network architecture takes care of (p.10):

Denial of service: Sometimes, legitimate users are not able to access their call service, due to some fault in network configuration.

Theft of service: Sometimes, a subscriber sets a VoIP account with false billing information that cannot be detected by network architecture.

Invasion of privacy: Like electricity and cable connections, network can be intercepted. Every subscriber has an individual cable to prevent this.

## 2.3 Security issues in p2p VoIP systems

With VoIP theft reaching $1 million in 2006, two arrests having been made so far, at least 15 p2p VOIP service companies (e.g. Skype, Verizon, etc.) have been intercepted, one of the companies having been hacked up to $300,000 (Waldron, p.9), security issues in p2p networks are of paramount importance, and the methodology of securing p2p systems, are the focus of this topic.

As per Skype method, a 256 K AES infrastructure device, is the only safe, secure and reliable means to deliver p2p VoIP services (Waldron, p.11). The architecture used for this purpose, is called Public Key Infrastructure (PKI). Since our case study is based on Skype, it is useful to discuss the code signing authority for this p2p networking client. In Skype's case, VeriSign, which is a market leading authority, provides PKI protection. VeriSign is used by some of the most illustrious online service clients, and its network is literally, like an impregnable fortress.

So, what does VeriSign basically do? It basically vets of, vouches for and binds public keys to users. This is done by carrying out software at a central location, with other coordinated software at distributed locations. The public keys are embedded in certificates, which are nothing but electronic algorithms to encrypt, and decrypt messages, traveling to and fro. A user can digitally sign messages, using his private key, and another user (Skype) can check that signature, using software packages (VeriSign, 2006).

As an example, here's a sneak peek into Skype's methodology to arrest a security violation, using PKI infrastructure, set in place (SkypeSecurity, 2006):

Problem description: A security bug in the Skype for user windows has been identified, and fixed.

Discussion: A defaulter, who connives a Skype URL, that is malformed in a particular way, has done it due to incorrect parsing of the parameters, by system vulnerability, especially an attack on the source code of the web page, and which leads to an incorrect file transfer.

Impact and affected software: All Skype releases, before 2.5 are affected due to the attack, and the impact can be fatal.

What does Skype do? The preferred method to install security updates, is by directly downloading from Skype website. For Skype with Windows OS environment, the digital certificate is signed by "VeriSign Class 3 Code Signing 2004 CA".

A security bulletin pop-up feature automatically comes up, with installation, and the threat is removed. It contains an array of base vectors which have a correct display scheme, pre-assigned by VeriSign, something as follows.

```
Access Vector (AV) ........... Remote
Access Complexity (AC) ....... Low
Authentication (Au) .....,.... Not Required
Confidentiality Impact (C) ... Partial
Integrity Impact (I) ......... None
Availability Impact (A) ...... None
Impact Bias (B) .............. Confidentiality
```

Computed CVSS base score: 3.5

## 2.4 Cross-layer security architecture (mobile communications)

When talking of VoIP services, mobile communications using wireless technology is an important issue. Unlike infrastructure-based networks, mobile communications suffer from severe performance problems due to their dynamic nature The medium is shared, and interference-prone, routes are unstable, energy can be a limiting factor for sensor nodes (Winter et al, p.1). To overcome these limiting barriers, cross-layer architecture are a promising alternative. They help by minimizing the impact of above problems, and increasing the scalability and reliability of networks in mobile communications (Winter et al, p.1).

As discussed earlier, in the chapter on enterprise and carrier layer architecture, layered architecture have the advantage of reduced design complexity, more modularity, and improved maintainability, compared to monolithic stacks (p.1). This modularity allows for combinations of different protocols, flexibility in terms of tracing an error back to a given layer, and improving the general performance of the network stack (p.1). The main problem with simple-layered architecture, lies in the fact that they are shared, unreliable mediums leading to bit errors, high collisions, high delays and lowered thoroughput (p.1. In contrast, cross-layering provides the unconventional comfort in these parameters, by sharing information among different layers, that can be used as input for algorithms. These processes of information sharing, are coordinated and structured somehow, the enhancement of the layered architecture is preserved, and architecture longevity guaranteed (p.2).

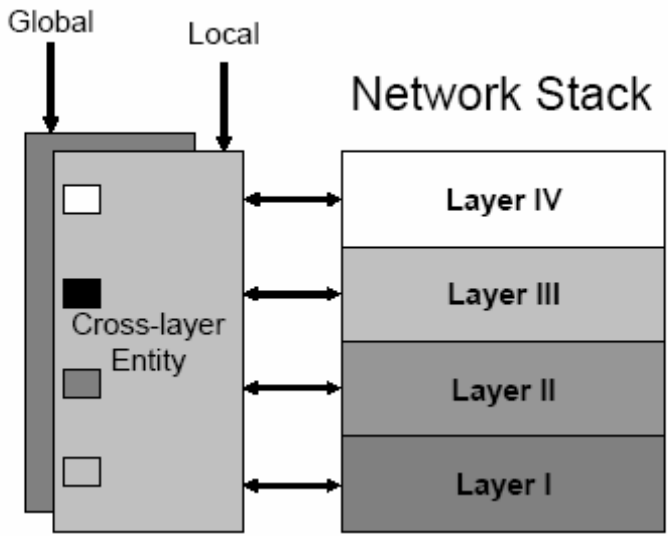

**Figure 4: Cross-layered architecture**

The cross-layered architecture, endeavors to maintain a network-wide, global view of multiple metrics, such as load balance, battery status, and routing destinations (p.2). In order to discuss simplistic designs of cross-layered architecture, we can consider suitable local view, global view considerations. Local view consists of specific nodes, that can be used for local *optimizations* such as load balance (p.3). The global view, tends to *propagate* the information, by collecting numerous samples of information from local nodes (read packets of information) (p.3)

Most mobile communications companies follow what is called *Common Data Security Architecture* to implement the cross-layered objectives, downloadable from http://www.opengroup.org/security/l2-cdsa.htm

## 2.5 Carrier peering and session border protocol

Both Carrier Peering and Session Border Control are important security issues for VoIP transactions. It is with this sense of importance, that the concepts should be made clear.

Carrier Peering refers to the ability to *interconnect* networks (Sonus, p.2). While it is possible for all carriers to interconnect their switching nodes directly, this is far too vague, and would not provide the kind of *control* desired from such inter-connectedness (p.2). Therefore, traffic is controlled by what are known as *inter-connects* or *peer points* (p.2). Peering is a complex process of interworking protocols and variants, as well as digital manipulation, parameter interpretation, screening, route prioritization and much more (p.2). While the specific requirements and protocols change, based on the peering point, and its location in the network topology, the primary role of peering device is to secure the carrier's networks, based on SIP protocol, the standard for VoIP communications (p.2).

Important facts to be kept in mind over carrier peering: 1)Back-to-back conversions through a pair of media gateways, an intrinsic ritual, that *raises* call costs, and also degrade performance by what is known as *packetization delay* (p.3) and 2)Signaling must be converted, from SIP or H.323 commonly used in VoIP protocols, to solve interworking problems (p.3).

Security Carrier Peering performs some other functions, desired in VoIP and mobile communications (p.3): 1)Billing, i.e to generate billing records when a call is completed 2)routing services (8YY translation, ring tones, announcements, and proportional routing).

Session Border Controls (SBC), also called Session Awareness Firewalls, are primarily security devices designed to protect the network, like any security firewall (p.4). In layman's terms, the functionality is crucial for a robust VoIP system, that can fend off attacks, by problems such as Denial of Service (DoS), and unauthorized interference by illegitimate users (p.4). *So, how does it do the screening*? SBC's basically provide a single, or small range of IP addresses, through which all media and signaling must pass (p.4), and other functions are similar to Carrier Peering, such as Billing, Routing services, Traffic flow management, etc. (p.4).

Generally, these two services are provided by Network Solution companies, such as Sonus, and Juniper, based in CA, USA. For economical purchase in Project decisions, it is useful to go for what is known as Juniper Networks VF-series Session Border Control Solutions, a comprehensive package that includes service assurance, hosted NAT traversal mechanism (more on this in next chapter), managed enterprise IP telephony, and most importantly, *legal formalities* (Juniper, p.3).

## 2.6 <u>Security issues with NAT traversal mechanism</u>

In earlier chapters, we discussed the importance of *routers* in infrastructure case studies. In real-life scenario, several other entities, called *middle-boxes* affect the quality of traversal of voice packets, by inserting the flow of transport protocols, or by translation. Some of these middle-boxes are common names: IP tunnel endpoints, markers, proxies, caches, transport relays, modified DNS servers, etc. (HIP Research Group, 2006). In layman's terms, these can be identified as *stumbling blocks* to efficient VoIP communications. So, this chapter basically deals with VoIP across Network Address Translators (NAT's), which is the more common terminology for the middle-boxes (HIP Research Group, 2006).

Although this subject runs very deep, it is possible to recommend the following 3 security activities that are intrinsic to traversal of NAT mechanisms (HIP Research Group, 2006):

- Opening pinholes in firewalls (i.e. loading firewall rules, allowing packets to traverse), and creating NAT bindings are highly security-sensitive actions.
- To reduce the complexity of the overall protocol, it's possible to use something called UDP encapsulation, which is used to correlate outgoing, and incoming signal values.
- Using NAT extensions

## 2.7 End user and hop-by-hop authentication techniques

In this chapter, we will mention a string of successful techniques, that are used in end user, and hop-by-hop authentication techniques (Computer Society, 2006):

- Longer addresses, 128 bits compared to 32 bits: As of today, there are 4 billion IP addresses, that will grow into 9 billion by 2050, not to men-tion the cross-layered architecture functions. That would lead to this demand.

- Simplification of headers: Processing power will increase
- New extensions to support Options, and Data Integrity
- Flow labeling capacity: Labeling of packets for regulating particular traffic flow
- Automatic configuration: For faster VoIP transfer, plug-n-play automatic configurations are desired.
- <u>Hop-by-hop</u>: This carries information, that must be processed, and carried by every node along a packet's traversal path.
- Data authentication, fragmentation and routing: this is as per discussions made in earlier chapters.

## 2.8  <u>Vulnerabilities in VOIP protocol, e.g. SIP</u>

In addition to providing telephony services, any VoIP device must provide for mechanisms to protect from toll frauds, eavesdropping and call-hijacking, among other things. (Networksystemdesign, p.1). SIP was designed only for the purpose of making communications possible. The SIP-based VoIP network is useful from the point-of-view of a layered approach (p.2). Threats, and therefore, countermeasures, can be mapped to the layers of the network-reference model (p.2). The defense strategy mainly consists of a three-layer security model, which goes as follows:

- <u>Infrastructure security layer</u>: Protect and secure the network architecture
- <u>Network services security layer</u>: Protect, and access end users.

- Application security layer: Protect and secure, SIP-based communications

Based on general security precepts, each security layer has to be evaluated based on the following premises (p.2):

- Authentication: Confirm the identity of communicating entities, whether individuals, entities, devices, services or applications, mainly using authentication guards.
- Authorization: Cross-checks identity for role and access. This provides unauthorized access to services, and identity frauds.
- Accountability/Audit: Keeps track of usages, and security services, and provides early intimidation on security threats, and also helps in recovery.
- Availability/Reliability: Redundancy, perimeter-protection, and hardening ensures that authorized users continue to have access to network services, despite DoS attacks, and other security hazards.
- Confidentiality: Encryption of communications streams, prevents unauthorized entry, with better access control.
- Integrity: Prevents unauthorized deletion and replication of data.
- Non-repudiation: A method to prove communications actually occurred.
- Privacy/Anonymity: Privacy addresses issues like phone number harvesting, cell pattern tracking, etc. that violates user privacy.

For private network systems companies, like Juniper, following provisions are made to ensure SIP Protocols are kept secure, and user-friendly (JuniperSIP, p.3):

- Stateful signature: Juniper network provides signatures, that protect against a number of exploits, that target Skype, and other VoIP vendors.
- Protocol anomalies: Recall the chapter covering how Skype fights hacker intrusions. That is a summation of SIP guides, to arrest protocol anomalies, which may vary a lot, as follows:

SIP Non-standard method: The signature detects a SIP request not defined. SIP Wrong version: This can get automatically corrected by proper download of Skype version (2.5 latest). SIP No colon after command. SIP Method Overflow. SIP unknown header. SIP Wrong content length. SIP No Protocol ID's.

## 2.9 QoS impact due to security implications

As per Research paper used of National Institute of Science and Technology (NIST), security steps discussed in the dissertation so far, cause an irreversible chain of impact on Quality of Service of VoIP systems. This is the presentation of findings of the paper (NIST, 2006):

- Latency: Latency refers to the time it takes for a voice signal to go from source to destination. Ideally, it should be zero, but there are upper and lower toleration limits (p.19). In the United States, acceptable limit is 150 ms for one-way traffic. For international calls, it is 400 ms (p.19). Any VoIP service provider must allow for a Latency budget, which may look like this (p.20):

| Delay Source (G.729) | On-net Budget (ms) |
|---|---|
| Device Sample Capture | 0.1 |
| Encoding Delay (Algorithmic Delay + Processing Delay) | 17.5 |
| Packetization/ Depacketization Delay | 20 |
| Move to Output Queue/Queue Delay | 0.5 |
| Access (up) Link Transmission Delay | 10 |
| Backbone Network Transmission Delay | Dnw |
| Access (down) Link Transmission Delay | 10 |
| Input Queue to Application | 0.5 |
| Jitter Buffer | 60 |
| Decoder Processing Delay | 2 |
| Device Playout Delay | 0.5 |
| Total | 121.1 + Dnw |

**Figure 5: Sample Latency budget**

- <u>Jitter</u>: Jitter refers to non-uniform packet delays (p.21). It is often caused by low bandwidth situations, and can be experimentally detrimental to QoS than actual delays. Jitter can cause packets to arrive, and be processed out of sequence. When jitter is high, packets arrive at their destinations in spurts, a situation that is analogous to several automobiles coming to a grinding halt at a traffic signal. There are basically three useful mechanisms to control jitter (p.21): 1)Usage of a buffer: It is used at the endpoint of traffic, where the buffer has to release its buffer packets once every 150 ms. 2)Routers with effective headers 3)Efficient use of bandwidth; a great deal of jitter is taken care of, by effective use of bandwidth.
- <u>Packet loss</u>: VoIP is excessively intolerant of packet loss, which forms a steady order indeed (p.21). Packet loss is a by-product of excessive latency, or jitter. Compounding the packet loss problem is, VoIP's reliance on RTP, which uses unreliable UDP for transfer. By the time a packet

> could get reported for something, OoS always gets exceeded. Packet
> loss is an unavoidable reality, and barring increasing bandwidth, there
> are no other independent solutions for counteracting packet loss (p.22).

Miscellaneous QoS security issues: SIP phone endpoints may freeze, and crash due to high rate of network traffic. SIP proxy servers may experience failure, with VoIP-specific signaling attack of under 1 MBPS (p.22).

## 2.10  Miscellaneous security issues

VoIP, being still at a nascent stage, security issues seem to be proliferating in response to its growth. Denial of Service (DoS) attacks against VoIP is a real possibility, as gathered by an interview of Frost and Sullivan analyst Jon Arnold (Kuchinkas, 2004). According to Arnold, each system requires a terminal adaptor node if it has to take care of the DoS problem. Another threat is Spam, which works similar to email threats. This affects the SIP code, and makes the language vulnerable to remote code executions and other cracks (Kuchinkas, 2004).

The majority of network operators use access-control lists, and destination-based BGP blackhole routing as the two primary methods of mitigating attacks (Pappalardo, 2006). According to the report, only 1.5% companies believe in reporting violations to law enforcement, mainly because of avoiding media pressure Security companies come up with suitable packages like Peakflow SP anti-DDoS software suite, that allows carriers to detect, and mitigate attacks (Pappalardo, 2006).

In order to boost network security, interconnections between nodes is supported by what is known as SS7 architecture (Signaling System 7), which is the signaling control system for the US and much the rest of world (Martinez, p.1). For Skype, VeriSign is the SS7 authority. The SS7 architecture consists of a high-speed, packet switch network, connected by three types of signaling links; Service Switching Points (SSP), Signal Transfer Points (STP), and Service Control Points (SCP) (p.1).

The use of data networks to transfer real time voice calls, is leading to proliferation in the demand of SS7 systems. Free VoIP to VoIP calls are made, best in a standard SS7 architecture (Martinez, p.2). Thus, for any signaling protection module, it is compulsory to include an SS7 architecture.

For wireless security issues, and for a Voice over Wireless LAN deployment, the first project goal to consider, is the high quality calls expected by users (Jacobs, 2006). Aruba, (www.arubanetworks.com) has a full range of solutions over this particular domain. Aruba's partners include some other big names, in the Wireless industry, Spectral Link, Quote, Chart, Avaya, voice-badge maker Vocera, and software designer TeleSym. The objective of securing VoWLAN, lies in proper MAC addresses authentication, and WEP analysis. In addition, radio frequency management (RF) and load balancing are the important functions of relevant VoWLAN security mechanisms (Wi-fi Planet, 2004). The Aruba 5000 switch is capable of handling 5000 calls simultaneously, mainly due to codecs and compression schemes used by each of its partners.

The problems with VoWLAN start in their setting up functions. In contrast to data over wireless networks, VoWLAN needs are much more complex. The basic task of Wi-fi, is to provide an interoperability testing for the new IEEE 802.11 equipment flooding the market. It is up to the technical task group in a Wi-fi Project, to allow for more expansion in its current, and future scope (Frank Bulk, 2005).

# Chapter 3. METHODOLOGY/SIMULATION SETUP

Based on literature sources discussed in previous chapters, we will summarize our knowledge on a conceptual design framework, for a sample VoIP Project.

Basic Components of VoIP Infrastructure: 1)Router devices 2)Switch devices 3)Load balancing devices (All LAN devices based on **Ch 2.2**) 4)Firewall devices 5)Virtual Private Networks (VPN's) (All security devices based on **Ch 2.2**) 6)Call Server 7)Gateway 8)MCU 9)Ethernet Switch 10)Routers (Other components)

As per (**Ch 2.5)**, it is prudent to arrange the components, as per cross-layered architecture, for enhanced load balancing, reduced design complexity, and higher modularity, and maintainability, therefore ensuring suitable packet speed, thus ensuring the design is ready for Mobile communications as well.

Type of VoIP Project: For our design purposes, we will examine p2p (peer to peer) VoIP communications, as per (**Ch 2.3)**, since our Simulation experiment will be based on Skype $^{TM}$, which is a p2p communications tool. Central to a p2p design set's security standards, is an 128-digit AES-based cryptographic algorithm (**Ch 2.1**), which can be downloaded from http://csrc.nist.gov/publications/ after e-payment. The security operates out of a Public Key Infrastructure platform, e.g. VeriSign $^{TM}$, which attests digital signatures to all VoIP calls being made (**Ch 2.3**). A p2p infrastructure also necessitates a standard network protocol, which the data packets can identify. As per (**Ch 2.1**), we must go for Sessions Initiation Protocol (SIP) which is the most widely-used network protocol. For effective protection against security breaches, the SIP protocol must be under the shield of some network guard, and Juniper $^{TM}$ (**Ch 2.8**) is a leading authority to provide digital signature authentication for secure SIP voice transactions. In order to smoothen voice quality for SIP protocol, there are some techniques to be kept in mind, called hop by hop authentication techniques (**Ch 2.7**).

Security framework: Carrier peering, and session border controls are effective instruments designed to improve upon current security protocols (**Ch 2.5**). These are provided by network solutions companies, like Sonus $^{TM}$ and Juniper $^{TM}$. SBC's provide a small range of IP addresses, through which all signal communications must pass. Both these operatives perform some other vital functions desired in VoIP communications 1)Billing, i.e. to generate billing records when a call is completed 2)routing services (8YY translation, ring tones, announcements, and proportional routing). 3)Caller ID etc. We also need appropriate mechanisms to perform NAT traversal functions, e.g. overcoming communication roadblocks like IP tunnel endpoints, markers, proxies, caches, transport relays, modified DNS servers, etc (**Ch 2.6**).

Improving QoS: Building reliable security infrastructure comes at the cost of declined satisfaction in Quality of Service (QoS) for VoIP communications. Here are some appropriate measures designed to improve upon this factor (**Ch 2.9**):
- Latency: Should be kept between 150 ms- 400 ms (for international calls)
- Jitter: Is regulated using routers with headers, or buffer instruments.

For best QoS, enough bandwidth is a must as proved in next page, and broadband networks should be encouraged. SS7 architecture, and VoWLAN networking are essential for a p2p broadband experience (**Ch 2.10**). Aruba [TM] is the biggest name for VoWLAN, whereas Juniper [TM] is used for designing SS7 architecture.

The main issue of concern in design is, *Voice Quality*. VoIP is a synchronous and real time application (Qovic, p.1). When it combines with IP, call quality problems may result (Qovic, p.1).

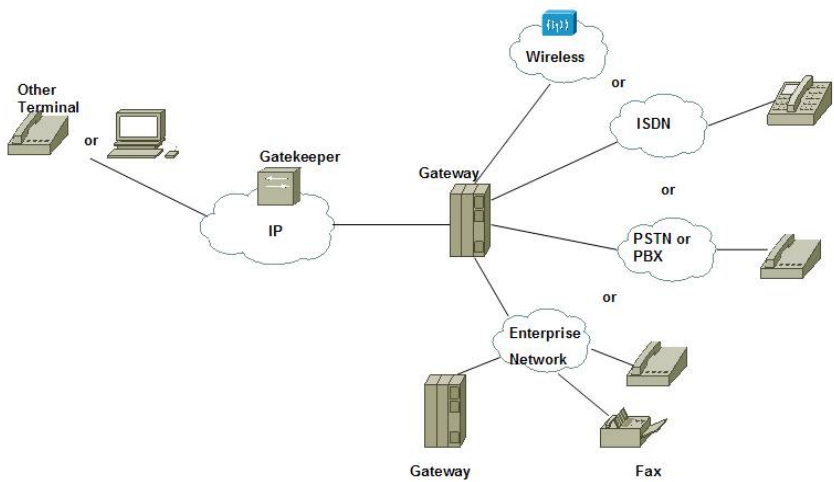

**Figure 6: Sample VOIP Network**

In considering the totality of VoIP design, we must examine the factors that negatively affect voice quality, and any fault tree analysis, must take a subjective account into the quality of service, being addressed as a central issue. Perceived voice quality is a function of many factors: jitter, delay, echo and packet loss (Qovia, p.4). Jitter and delay have been covered in a previous paragraph. Echo, is caused when hybrid circuits in a telephone network convert between a four-wire circuit, and a two-wire circuit (p.4). Echoing occurs when there is an audible leak between sending and receiving areas. Echo is reduced by physically shortening the distance between the sending and receiving circuits (p.4). In order to quantitatively portray the performance of VOIP modules, it is important to ponder over QoS issues, the following performance table is useful (Qovia, p.5):

| Parameter | SCORE | | | | |
|---|---|---|---|---|---|
| | 1 | 2 | 3 | 4 | 5 |
| Latency (ms) | < 50 | 50 - 75 | 75 - 100 | 100 - 200 | > 200 |
| Jitter (ms) | < 5 | 5 - 10 | 10 - 50 | 50 - 100 | > 100 |
| Packet/Loss (%) | 0 | 0 - 1 | 1 - 2 | 2 - 3 | > 3 |

**Table 2 : Voice Quality Measurement**

VoIP deployment: Table 2 will be of great help in the course of Project Simulation, in next chapter. Meanwhile, central to design is consideration of proper deployment of VoIP. Deployment of new products has always been a great challenge, right from technology officers, to network administrators, They have to balance productivity, with costs, and integrate it with existing infrastructure (Rao, 2004). Other issues to deserve priority, are *troubleshooting*, *teething* and *compatibility* problems during initial stage, and *performance tuning* and *maintenance* later. So, here we discuss different parameters connected with a reliable p2p VoIP deployment (Rao, 2004).

- Bandwidth: As voice is more sensitive than data, bandwidth requirements automatically go up. VoIP devices should be configured to use a more bandwidth-intensive codec, such as G.729-A, which can use 87.2 kbps of NEB (Nominal Ethernet Bandwidth). Also, existing traffic patterns on the network should be identified. If it is already congested with a lot of broadcast traffic, it would definitely affect response time. For best results in p2p communications, it is safer not to merge them with other signals, such as TV and Radio. When deploying VoIP, it is mandatory to have a separate VLAN, which will keep the voice and data networks separate (Juniper's VoWLAN in our case). Coming back to bandwidth, the important design criteria is to ascertain the number of simultaneous voice calls that can be made over WAN links (**Ch 2.2**). For this, compression technique to be used should be known, the payload size of voice packets. Some VoIP calculators are available, e.g. in Project Simulation chapter. www.voip-calculator and www.ixiacom.com are VoIP deployment modules available online.

- QoS: Typical toll-quality calls, require at least 16-20 kbps of bandwidth. There are also bandwidth management solutions (**Ch 2.9**).

- Handling power problems: Unlike regular phones, p2p phones suffer from power outrages. However, a good-built UPS is an effective solution. However, a good bit of power planning from source end is desired, as to how much backup power is needed.

- Security issues: Apart from security mechanisms discussed in the dissertation, the network should be subject to routine maintenance, taking regular back-ups, installing anti-virus software, etc. Also, implementing VoIP performance analyzers, like "Appmanger 6.0" from NetIQ for smooth functioning of VoIP is a good idea.

- Prerequisites before deployment: 1)At each LAN, network architecture should be such that at least 70% throughput is achieved within 10 msec.

2)Round the trip delay of each packet of IP should be no more than 10 msec. 3)Round the trip delay of each packet in LAN and WAN together should not be more than 150 msec.

Thus, we discussed the feasibility of a proposed VoIP design, and the enormous importance to be placed on security installations. Next chapter will deal excessively with a safe and secure VoIP software, (Skype), and the strain put by its security paradigm on QoS parameters, e.g. Jitter and Latency.

## 3.1 VOIP project simulation

Throughout this dissertation, a consistent attempt has been made to study VoIP technology with the aim of working out a feasible Project on an actual scenario. While the previous chapter spelled out details regarding elementary design considerations, this chapter will undertake a Simulation exercise around a real VoIP environment, based on a real p2p VoIP software. For our simulation purposes, we will consider Skype. The previous chapter dealt with design background of Skype. Also, in earlier chapters, we dealt with security mechanisms, a great part of which dealt with p2p VoIP software. Before proceeding ahead, it is important to recapture the following essential points of Skype software:

- Skype uses p2p network to overcome the barriers of firewall and NAT traversal mechanisms, using routes. The p2p model is different from the client-server model discussed in (**Ch 2.2**) and (**Ch 2.4**) by the fact, that the user directory is decentralized and distributed among nodes in the network, with the end-user nodes (called supernodes) having higher bandwidth, for increasing clarity. The Skype network can scale a user base of 100 million, without the need of a centralized network.

- Skype's code is closed source, using proprietary SIP protocol (**Ch 2.8**).

- For security procedures, Skype uses 128-bit AES cryptographic algorithm devised by VeriSign (**Ch 2.3**), and users have to log in using their unique ID's and passwords, and is one of the safest in industry. Skype's window rooms makes extensive use of Carrier Networking and Session Border Controls (SBC's) (**Ch 2.5**).

For Simulation purposes, we will make real calls using Skype and test its impact on QoS parameters, as well as Lines and Bandwidth requirement for actual field condition, using a VoIP calculator, i.e. for a given packet duration, the number of voice paths permitted for any range of bandwidth. Using data as suggested, we can perform an independent statistical analysis too.

*VisualRoute2006*, http://www.visualroute.com/index.html is a software that enables determining whether a connectivity problem is due to ISP, the Internet or the Web site being visited, and pinpoints to network where the *fault* lies. We have used VisualRoute Business edition for checking Skype calls. These are important observations for connectivity with Skype from host server.

| Hop | %loss | IP Address | Location | Tzone | ms | Network |
|---|---|---|---|---|---|---|
|  |  | \| 62.216.129.146 \| | London, UK | ***** | 308 | FLAG Telecom |
|  |  | \| 129.250.10.201 \| | Centennial, CO | -7 | 313 | FLAG Telecom |
| 2 | 10 | \| 129.250.2.38 \| | Centennial, CO | -7 | 317 | \| NTT America, Inc. NTTA-129-250 \| |
|  |  | \| 129.250.2.32 \| | Centennial, CO | -7 | 320 | \| NTT America, Inc. NTTA-129-250 \| |
| 4 | 10 | \| 129.250.2.122 \| | Centennial, CO | -7 | 320 | \| NTT America, Inc. NTTA-129-250 \| |
| 5 | 10 | \| 129.250.2.237 \| | Centennial, CO | -7 | 317 | \| NTT America, Inc. NTTA-129-250 \| |
| 6 | 10 | \| 129.250.16.11 \| | Centennial, CO | -7 | 307 | \| NTT America, Inc. NTTA-129-250 \| |
| 7 | 10 | \| 129.250.223.66 \| | Centennial, CO | -7 | 311 | \| NTT America, Inc. NTTA-129-250 \| |
| 8 | 10 | \| 198.173.5.35 \| | Centennial, CO | -7 | 299 | \| NTT America, Inc. NTTA-129-250 \| |

**Table 3 : Observations for connectivity with Skype from host server**

As per above observations, the call was made from London, UK to Skype server in Centennial, CO, USA. Note the number of routings (7 et al) utilized before the call connects to main server. As discussed earlier, Skype utilizes these routings for NAT traversal, as well as avoiding firewalls. Also, Skype uses End user hop-by-hop authentication techniques (**Ch 2.7**), in this case we have 8 hops between London and destination, meaning the call was almost *instantaneous*. The roundtrip time (latency) to Skype.com, as shown in table, is 299 ms, which is a third of a second. Regarding QoS parameters, packet loss never exceeds 10% as per **Table 2**, it falls above 200 ms, which is defined as a fairly *stable* zone for International calls (It should be between 150-400 ms). The jitter is at the most, 100 ms.

# Chapter 4. RESULTS

The aim of this simulation, was to show Skype is a reliable VoIP mechanism, despite having a credible security infrastructure in place. Indeed, Routing is a very significant method of improving end-user Quality of Service. Considering the security infrastructure laid for Skype calls, the model may be replicated for any Project on VoIP. Next in important scheme of things, lies another round of simulation. But, this time we shall utilize the services of what is called VoIP-calculator. http://www.voip-calculator.com/calculator/. There are plenty of parameters within which a VoIP calculator operates, but we will focus our attention on only one, the Erlangs and IP Bandwidth calculator, which quantitatively estimates the number of bandwidths required through an IP based network for a fixed number of voice paths. Using statistical analysis, we will determine if the bandwidth design fulfills a statistical fit of sorts.

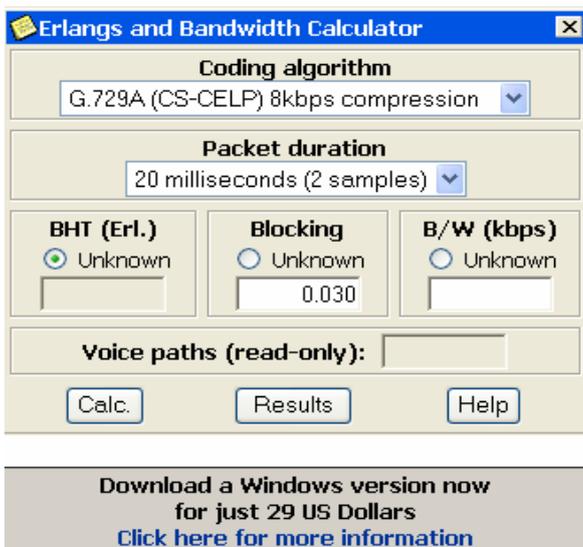

**Figure 7 : VoIP Calculator**

As per Figure 7, G.729A is the most common codec compression scheme (**Ch 2.1**). The frequency with which voice packets are transmitted, have a significant bearing on the bandwidth required. The selection of the packet duration is a compromise between bandwidth and quality. Lower duration requires more bandwidth, however higher duration means packet loss. So, where do we draw the line. For this reason, we test different calculations for different ranges of packet duration, and arrive at a good ballpark figure, using Statistical means.

A little explanation on different calculator parameters. BHT refers to Busy Hour Traffic, which refers to the number of hours of call traffic during the busiest hour of operation in a telephone system. Blocking refers to the failure of calls due to an insufficient number of calls available. Typically, a figure such as 0.03 would mean that 3 calls out of 100 attempted, have been blocked. Bandwidth is the amount of bandwidth required in Kbps to run the traffic. For optimum compromise between duration, and packet loss, our calculations proceed at packet duration of 30 ms (3 samples). Considering bandwidth is fixed, at Broadband speed of 120 Kbps, we concentrate on other parameters and prepare the table as below. Our ultimate objective is to fine-tune a solution between no. of hours allowed on busy traffic, versus no. of calls dropped in the process.

| Busy Hour Traffic | Calls dropped |
|---|---|
| 1.900 | 0.01 |
| 2.250 | 0.02 |
| 2.500 | 0.03 |
| 2.750 | 0.04 |
| 2.950 | 0.05 |
| 3.100 | 0.06 |
| 3.300 | 0.07 |
| 3.450 | 0.08 |
| 3.600 | 0.09 |
| 3.750 | 0.1 |

**Table 4 : VoIP calculator result at 120 Kbps bandwidth, and 30 ms packet duration**

For analysis of above Simulation, we will utilize a Statistical software called Minitab 14.0 (www.minitab.com) Minitab is a market leader in the field of Statistics. As appears above, there seems to be a lot of correlation between two variables above. Statistically, we verify this claim, and indeed it is true, as the Pearson Correlation coefficient of Busy Hour traffic and calls dropped, comes at 0.991 or 99.1% (Refer Appendix B).

To determine other statistical values, and gather inferences, we perform a sample regression analysis, with Busy Hour Traffic as Predictor (X) variable, and Response (Y) variable (Refer Appendix C). As per regression analysis done, we have the following results:

*Calls dropped = - 0.0915 + 0.0496 x Busy Hour traffic*

Also, R-squared value at S-value, comes at 98.1%, which is again a high value. Thus, both correlation and regression analysis results, point to the fact that the more we clear up network lines to accommodate busy traffic, it also increases proportionately the number of calls dropped, due to obvious increase in packet loss rate due to increased traffic.

In order to statistically determine the optimum number of calls that may be dropped, for achieving our bandwidth versus quality compromise, we perform a Moving average time series test clearly, points 1, 2, 3, 4, 9, 10 fail the test and should not be considered for design criteria.

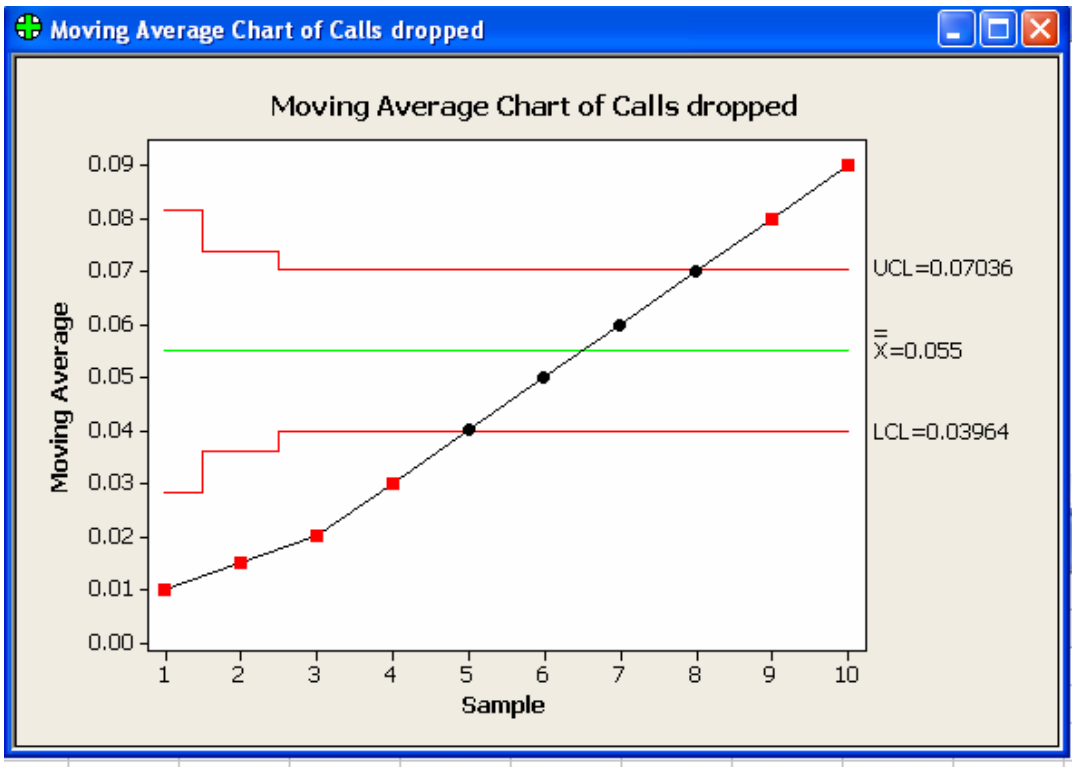

**Figure 8 : Moving Average Cell Chart**

Thus, for safe design, we can consider any of the points 5, 6, 7, 8 and based on central tendency theorem, we therefore opt for the following design values:

| Bandwidth | 120 Kbps |
|---|---|
| Packet loss | 10% |
| packet duration | 30 ms (3 samples) |
| Coding algorithm | G-729 A (8-bit compression) |
| Busy Hour Traffic | 3hours |
| No. of calls dropped | 0.052 |

**Table 5 : Final Design Values for VoIP Project Simulation**

Of course with increase in bandwidth, the value "no. of calls dropped" decreases proportionately, and broadband connections are therefore, encouraged for VoIP applications.

# Chapter 5. CONCLUSION AND FUTURE WORK

## 5.1 Lessons learned

The major achievements in this dissertation are summarized as follows:

- A blow-by-blow account of every single domain of knowledge that the field of VoIP incorporates; the basics of convergence, inputs from agencies connected in different applications of VoIP, e.g. VeriSign, Juniper, NIST, etc., major security initiatives, actual field measurement conditions, Quality of Service issues on security platforms. Design and Project Simulation is based on a strong, yet concise, theoretical background. It is no understatement to point out that the dissertation has managed to cover ground, on all desired areas of VoIP, drawn heavily from sound academic sources, and is a *user-friendly* manual for further study, or Project reference.
- A major achievement of this dissertation is, that it is developed along the lines of 13 security questions, gathered from expert agreement (**Ch 2.1**). These, in turn, form the backbone of Project Design (**Ch 2.2**), as well as Project Simulation (**Ch 3.1**). Almost all intermediate chapters have been presented in such a manner, that they support the objectives of the Project.
- In each chapter dedicated to understanding the particular security issue in question, important solutions are highlighted, to draw learning lessons.

- Enterprise and carrier network architecture, featuring routing devices, load balancing devices, switches, firewalls, VPN's (**Ch 2.2**), media servers, gateways, bandwidth managers, edge routers are the first tier of scalable network architecture, to be used in laying down VoIP infrastructure. The importance of Routing devices in reducing Round trip time for making calls, has been *proved* in VisualRoute Simulation experiment (**Ch 3.1**). For enhanced scalability, cross-layered architecture is a very strong basis for improving on factors, such as load balance, better firewall alert, and flexible algorithms (**Ch 2.4**).

- This dissertation has extensively looked into the phenomenon of peer to peer p2p VoIP communications (**Ch 2.3**), as opposed to client and server architecture. These operate on a standard instant messaging interface, and are protected using Public Key Infrastructure encryptions. VeriSign, is a leading authority in attesting digital signature to p2p services, like Skype.

- In order to register smooth functioning of VoIP networks, and introduce advanced features, such as call billing, caller ID, translation, ring tones, announcements, etc., many VoIP agencies (not Skype) make use of interconnecting nodes in their networks, called carrier peering (**Ch 2.5**), Session Border Controls (SBC's) are used for filtering of a narrow range of IP addresses, and prevent three main security hazards from happening: 1)Denial of Service 2)Theft of network 3)Invasion of Privacy (**Ch 2.5**).

- Speed is the most significant aspect of VoIP communications. Very few networks can afford delays due to congestion of lines, and providing enough bandwidth, is often not the solution. Many VoIP networks have to wind their way around intermediate middle-boxes, called NAT (**Ch 2.6**), and from end-user side, there is a necessity for hopping networks. Some hop techniques have been discussed in (**Ch 2.7**). In VisualRoute simulation experiment (**Ch 3.1**), usage of hops is depicted as a standalone function of improving upon inefficiencies creeping in, into the network.

- This dissertation takes a look into the impact of security infrastructure, on several QoS parameters; latency, jitter, echo and packet loss (**Ch 2.9**). Latency is an unavoidable reality in communications, and it can be kept within specification limits (150-400 ms), whereas jitter is non-uniform delay in arrival of packets, and it can be fixed using buffers, router headers and efficient use of bandwidth. (**Ch 3.1**) simulation deals with this area.

- Like any other communication media, VoIP has its own language protocol, Sessions Initiation Protocol (SIP), although H.323 and other languages are also used. The protocol is subjected to security hazards, and thus, requires to be vouchsafed by an authentication agency, e.g. Juniper Network. The protocol suffers from its own disadvantages, in terms of authorization, authentication and accountability.

- Among other security issues, the role of SS7 architecture, and VoWLAN have been detailed in (**Ch 2.10**). In order to boost network security, interconnections between nodes is supported by what is known as SS7 architecture (Signaling System 7). The SS7 architecture consists of a high-speed, packet switch network, connected by three types of signaling links; Service Switching Points (SSP), Signal Transfer Points (STP), and Service Control Points (SCP). VoWLAN stands for Voice over Wireless LAN feature, and is applicable in Project Design.

- With all available data at our disposal, the chapter on Design (**Ch 3**) collated a sum total of learning lessons for the VoIP project, and a conceptual framework on a cost-effective, and broadband quality VoIP design, and the essential parameters that would be needed to support the learning. The main themes of design stem from an eclectic combination of the best ideas discussed in previous chapters, and a basic deployment and implementation scheme, and the way the scheme should be carried out. The deployment incorporates the following parameters: troubleshooting, teething and compatibility problems during initial stage, and performance tuning and maintenance later. It identifies G.729 A as the most important bandwidth codec, and 87.2 Kbps as minimum bandwidth needs for reliable VoIP communications, 16-20 Kbps for QoS misalignments, UPS for handling power problems at source, and some elementary prerequisites to be taken care of, before the design can be open to audit. The main purpose of the chapter on design, has been to present a background for next chapter needs.

- The chapter on Project Simulation, (**Ch 3.1**), is the single most achievement of this dissertation, as it applies core principles gathered from study of the remaining dissertation. As different from the chapter on design, this chapter examines how and why QoS parameters and bandwidth metrics play a role in security-related prerequisites, and must be given due credit. In order to perform real life simulation, a real VoIP service provider was tested for QoS metrics. Skype was chosen because it is currently, the market leader in p2p VoIP communications, and reports drawn from Skype, can be helpful in ascertaining Project benchmarks. The first test consisted of ascertaining QoS parameters during a call made from London, UK to Skype server in CO, USA. The results for latency (round trip time), packet loss, jitter come well within established control limits, and the Simulation experiment serves as an excellent example, of how theory can be applied in practice. During simulation, special mention is made to the fact that all security features adopted by Skype for international calls, are documented.

Apart from QoS, there is a necessity to project the importance of bandwidth studies. For this simulation, we utilize an agency called VoIP Calculator, which quantitatively estimates the total bandwidth required for a given number of voice paths, provided other parameters remain same. The parameters we test in this simulation, include Busy Hour Traffic (BHT) against the number of calls dropped, other parameters remaining same. As evinced from calculator results, there is almost a linear correlation between the two variables, a fact that is confirmed later by Statistical analysis using both correlation, and regression methods. Statistical analysis is also used to find an optimized value for number of calls dropped, and the combined results, give us a safe set of design values, which is mentioned in table

## 5.2 **Future activity**

Since risk assessment on any VoIP project has carried itself off, for around six months, the future course of activity would lie in carrying out an actual "simulation" for the Project, and not just computer-based results, which comes only as an exemplary model. It is important to realize how current systems affect field work, and what new systems can be utilized. A little bit on futuristic systems is talked about in (**Ch 2.2**). Since simulation would be based on proposed methods as spelled out in Design, and lessons learned, the actual parameters will come up during field work for the same. The main functions included would be preparation of environment, server, equipment, software installations and configurations. VoIP is a world-standard application, and in order to meet world-class quality requirements, through customer feedback. Actual feedback will be based on fine tuning the system for Project details.

Here is how the proposed Project stands to benefit from the research undertaken for this dissertation:

- In terms of proprietary affiliations, the research looks into several agencies connected with VoIP projects; for instance, VeriSign for PKI attestation and SS7 architecture, Juniper and Sonus for Carrier Networking infrastructure, FIPS for AES algorithms, Aruba for VoWLAN installation. Apart from infrastructure companies, we also are introduced to testing affiliates, e.g. VisualRoute, and VoIP-calculator in Simulation module.
- Since SIP will be extensively used in any future Project development, the research has brought some insightful ideas into the fore, regarding the fault areas of that protocol, especially from a security angle. Some other agencies have been mentioned in relation to SIP interfaces (**Ch 2.8**).
- The chapter on design, (**Ch 3**) covers a featured deployment of VoIP projects, and is useful from the point of view of resource allocation. Special emphasis has been laid in the chapter on the indispensable nature of ele-mentary details, such as bandwidth requirements, QoS framework and security issues mandatory before deployment.
- The Chapter on Simulation, (**Ch 3.1**) goes one step ahead in consolidating our knowledge on VoIP projects, as it focuses on actual test practice results.

## 5.3 Prospects for future work

In order to consolidate our total knowledge gains, the research can benefit from complementary modules in other areas, that can aid in the development of VoIP Projects. Laying a VoIP infrastructure requires study of several aspects, closely connected to Project Development. Apart from theoretical foundation of a strong research base, it requires extensive study of Business development, particularly Project Implementation courses. Next in importance lies a core foundation of Legal areas; VoIP technology has came in for much criticism in many countries, e.g. China, where government-owned telecom agencies filed lawsuits against service providers such as Skype, and refused to cooperate because VoIP agencies are able to afford long-distance calls at much cheaper rates (Skype charges $0.021 per minute for any landline calls in the USA and EU). Also, there are concerns over Privacy laws in many countries. Although, Skype is a market exception, there are several VoIP service providers, who've simply jumped on the bandwagon with the boom in this technology spectrum, and a lot of improvement is desired, in terms of providing quality service, within legal framework.

Within technology spectrum, more research should be encouraged in mobile communications, and wireless (Wi-fi) technology, as these are integral to final phase of VoIP, and are vital, from a consumer point of view.

# REFRENCES CITED

# APPENDICES

## Appendix A. VISUALROUTE SOFTWARE MAIN INTERFACE

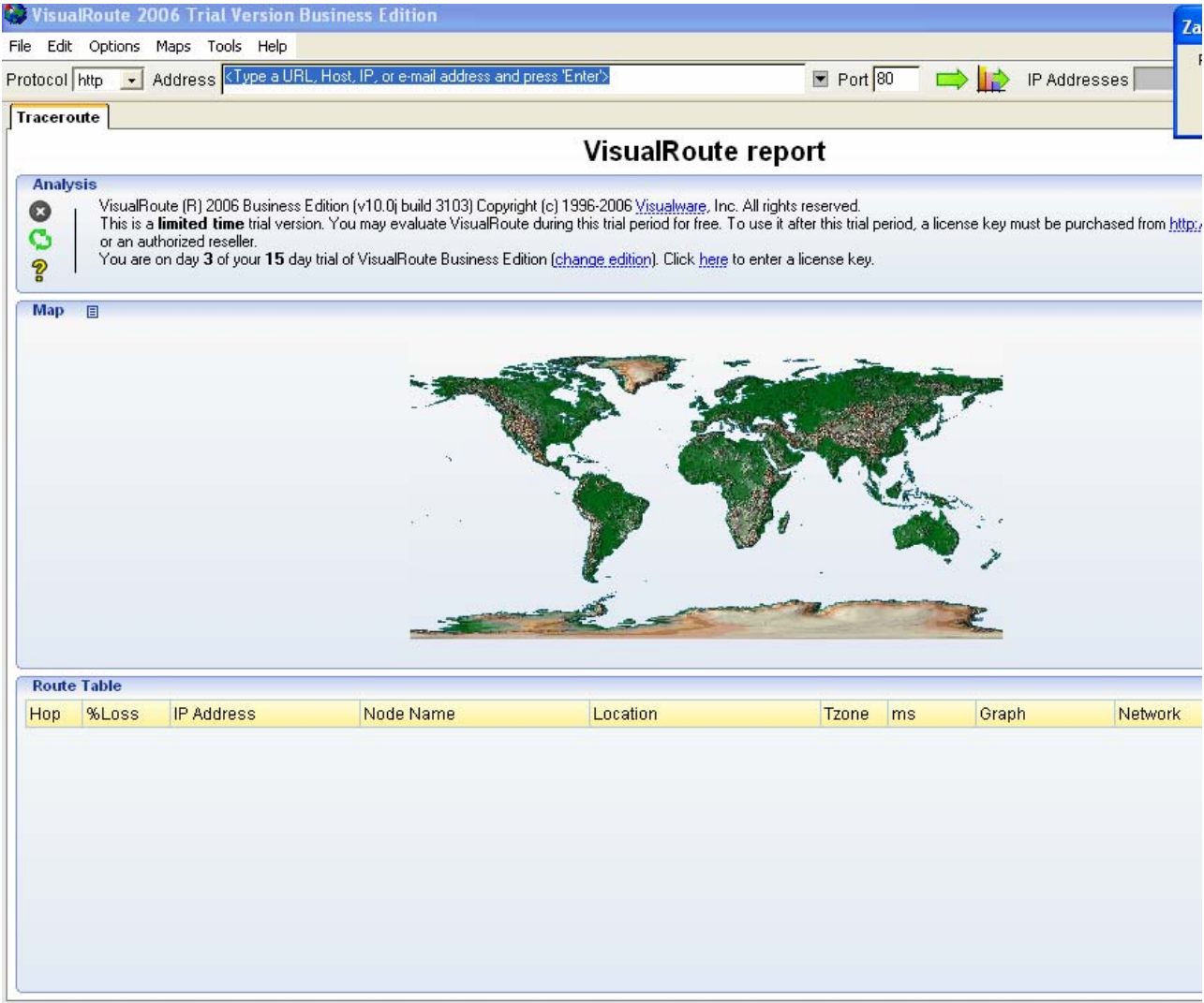

# Appendix B. MINITAB DATA PRESENTATION

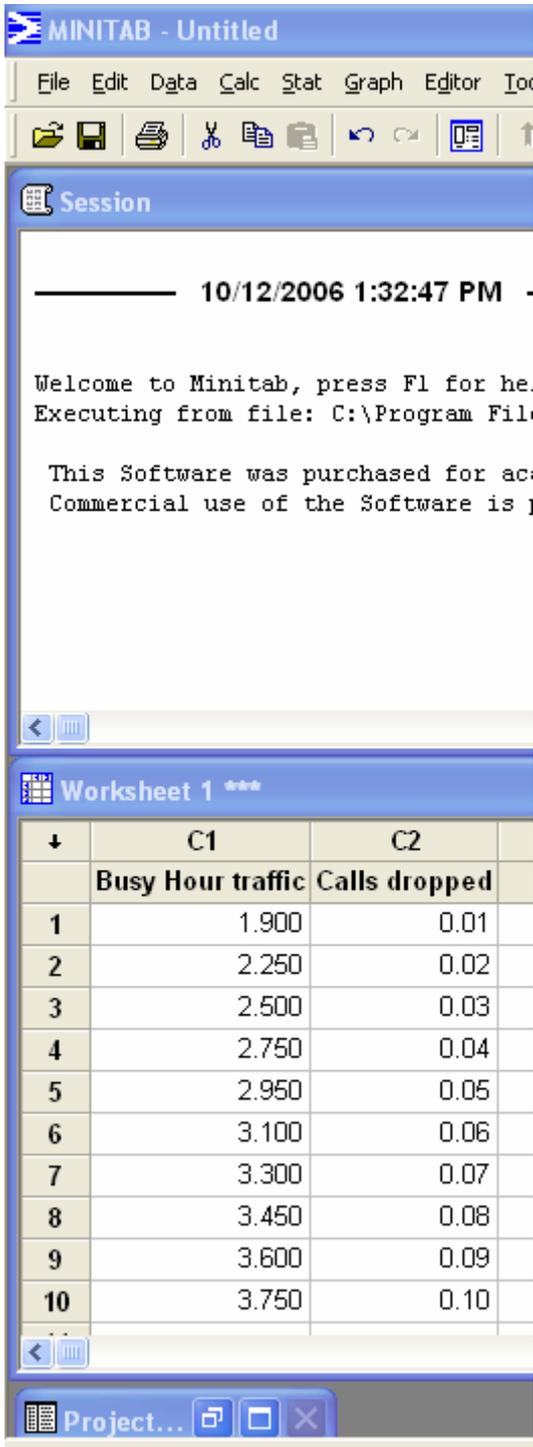

**Preliminary Data Presented in Minitab**

[Minitab Correlation dialog screenshot with Session window showing:]

10/12/2006 1:32:47 P

**Correlations: Busy Hour traffic,**

Pearson correlation of Busy Hour
P-Value = 0.000

Worksheet 1 data:

| # | C1 Busy Hour traffic | C2 Calls dropped |
|---|---|---|
| 1 | 1.900 | 0.0 |
| 2 | 2.250 | 0.0 |
| 3 | 2.500 | 0.03 |
| 4 | 2.750 | 0.04 |
| 5 | 2.950 | 0.05 |
| 6 | 3.100 | 0.06 |
| 7 | 3.300 | 0.07 |
| 8 | 3.450 | 0.08 |
| 9 | 3.600 | 0.09 |
| 10 | 3.750 | 0.10 |

**Performance of Correlation Analysis**

```
MINITAB - Untitled
File Edit Data Calc Stat Graph Editor Tools Window Help

Session
——————— 10/12/2006 1:32:47 PM ———————

Correlations: Busy Hour traffic, Calls dropped

Pearson correlation of Busy Hour traffic and Calls dropped = 0.991
P-Value = 0.000

Worksheet 1 ***
```

| | C1 | C2 | C3 | C4 | C5 | C6 |
|---|---|---|---|---|---|---|
| | Busy Hour traffic | Calls dropped | | | | |
| 1 | 1.900 | 0.01 | | | | |
| 2 | 2.250 | 0.02 | | | | |
| 3 | 2.500 | 0.03 | | | | |
| 4 | 2.750 | 0.04 | | | | |
| 5 | 2.950 | 0.05 | | | | |
| 6 | 3.100 | 0.06 | | | | |
| 7 | 3.300 | 0.07 | | | | |
| 8 | 3.450 | 0.08 | | | | |
| 9 | 3.600 | 0.09 | | | | |
| 10 | 3.750 | 0.10 | | | | |

**Pearson Coefficient Result**

# Appendix C. REGRESSION ANALYSIS REPORT

```
MINITAB - Untitled - [Session]
 File  Edit  Data  Calc  Stat  Graph  Editor  Tools  Window  Help

The regression equation is
Calls dropped = - 0.0915 + 0.0496 Busy Hour traffic

Predictor              Coef    SE Coef       T       P
Constant          -0.091527   0.007304  -12.53   0.000
Busy Hour traffic  0.049586   0.002426   20.44   0.000

S = 0.00440257    R-Sq = 98.1%   R-Sq(adj) = 97.9%

Analysis of Variance

Source          DF         SS         MS        F       P
Regression       1  0.0080949  0.0080949   417.64   0.000
Residual Error   8  0.0001551  0.0000194
Total            9  0.0082500

Unusual Observations

        Busy
        Hour    Calls
Obs  traffic  dropped      Fit   SE Fit  Residual  St Resid
  1     1.90  0.01000  0.00269  0.00291   0.00731      2.22R

R denotes an observation with a large standardized residual.
```

**Regression Analysis Report**

# Appendix D. MOVING AVERAGE CHART OF CALLS DROPPED

```
MINITAB - Untitled
File Edit Data Calc Stat Graph Editor Tools Window Help

Session
  Commercial use of the Software is prohibited.

Moving Average Chart of Calls dropped

Test Results for Moving Average Chart of Calls dropped

TEST. One point beyond control limits.
Test Failed at points:  1, 2, 3, 4, 9, 10

* WARNING * If graph is updated with new data, the results above may no
          * longer be correct.
```

| | C1 | C2 | C3 | C4 | C5 | C6 |
|---|---|---|---|---|---|---|
| | Busy Hour traffic | Calls dropped | | | | |
| 1 | 1.900 | 0.01 | | | | |
| 2 | 2.250 | 0.02 | | | | |
| 3 | 2.500 | 0.03 | | | | |
| 4 | 2.750 | 0.04 | | | | |
| 5 | 2.950 | 0.05 | | | | |
| 6 | 3.100 | 0.06 | | | | |
| 7 | 3.300 | 0.07 | | | | |
| 8 | 3.450 | 0.08 | | | | |
| 9 | 3.600 | 0.09 | | | | |
| 10 | 3.750 | 0.10 | | | | |

**Moving Average Chart Cells**